\documentclass[10pt]{article}
\usepackage[utf8]{inputenc}
\usepackage[T1]{fontenc}
\usepackage{amsmath}       
\usepackage{amssymb}       
\usepackage{bbm}
\usepackage{booktabs}      
\usepackage{multirow}      
\usepackage{siunitx}       
\usepackage{graphicx}      
\usepackage{caption}       
\usepackage[export]{adjustbox}
\usepackage[flushleft]{threeparttable}
\usepackage{subcaption}    
\usepackage{authblk}

\begin{document}

\title{AMO-ENE: Attention-based Multi-Omics fusion model for outcome prediction in Extra Nodal Extension and HPV-associated oropharyngeal cancer}

\author[1,2]{Gautier Hénique}
\author[1,2]{William Le}
\author[2,3]{Gabriel Dayan MD}
\author[1]{Coralie Brodeur}
\author[3]{Kristoff Nelson MD}
\author[3]{Apostolos Christopoulos MSc, MD}
\author[3]{Edith Filion MD}
\author[3]{Phuc-Felix Nguyen-Tan MD}
\author[2,3]{Laurent Letourneau-Guillon MD, MSc} %
\author[2,3]{Houda Bahig MD, PhD}
\author[1,2,*]{Samuel Kadoury PhD}

\affil[1]{MedICAL Laboratory, Polytechnique Montréeal, 
Montreal, Canada}
\affil[2]{Centre de recherche du CHUM (CRCHUM), Montréal, Canada}
\affil[3]{Centre Hospitalier de l’Université de Montréal (CHUM), Montreal Canada}
\affil[*]{Corresponding author: samuel.kadoury@polymtl.ca}


\maketitle

\begin{abstract}
    Extranodal extension (ENE) is an emerging prognostic factor in human papillomavirus (HPV)-associated oropharyngeal cancer (OPC), although it is currently omitted as a clinical staging criteria.
    Recent works have advocated for the inclusion of iENE as a prognostic marker in HPV-positive OPC staging.
    However, several practical limitations continue to hinder its clinical integration, including inconsistencies in segmentation, low contrast in the periphery of metastatic lymph nodes on CT imaging, and laborious manual annotations.
    To address these limitations, we propose a fully automated end-to-end pipeline that uses computed tomography (CT) images with clinical data to assess the status of nodal ENE and predict treatment outcomes.
    Our approach includes a hierarchical 3D semi-supervised segmentation model designed to detect and delineate relevant iENE from radiotherapy planning CT scans.
    From these segmentations, a set of radiomics and deep features are extracted to train an imaging-detected ENE grading classifier.
    The predicted ENE status is then evaluated for its prognostic value and compared with existing staging criteria.
    Furthermore, we integrate these nodal features with primary tumor characteristics in a multimodal, attention-based outcome prediction model, providing a dynamic framework for outcome prediction.
    Our method is validated in an internal cohort of 397 HPV-positive OPC patients treated with radiation therapy or chemoradiotherapy between 2009 and 2020.
    The segmentation model achieved a mean Dice score of 78.4 ($\pm$ 7.5), which improved to $83.5$ ($\pm 4.1$) following a component selection method.
    The ENE classification model reached $81.6\%$ ($\pm$ 5.7) in AUC for general iENE detection, and 89.9$\%$ ($\pm$ 5.0) when identifying high-grade (grade 3) ENE.
    For outcome prediction at the 2-year mark, our pipeline surpassed  baseline models with 88.2$\%$ ($\pm$ 4.8) in AUC for metastatic recurrence, 79.2$\%$ ($\pm$ 7.4) for overall survival, and 78.1$\%$ ($\pm$ 8.6) for disease-free survival.
    We also obtain a concordance index of 83.3$\%$ ($\pm$ 6.5) for metastatic recurrence, 71.3$\%$ ($\pm$ 8.9) for overall survival, and 70.0$\%$ ($\pm$ 8.1) for disease-free survival, making it feasible for clinical decision making. 
\end{abstract}

\flushbottom

\thispagestyle{empty}

\section{Introduction}\label{Introduction}
    
        Extranodal extension (ENE) in oropharyngeal carcinoma (OPC) refers to the spread of malignant cells beyond the capsule of metastatic lymph nodes \cite{AJCC8TH}, and is an indicator of tumor aggressiveness, as well as of a greater risk for invasion into adjacent tissues. Traditionally, ENE is confirmed histopathologically (pENE) through detailed examination following lymph node dissection, a time-intensive process. Presence of ENE is associated with escalated therapeutic regimens, including intensification of adjuvant and chemo-radiation protocols, as well as closer surveillance protocols \cite{hiyama_extra-nodal_2020}, which, while aiming to curb disease progression, also increases the burden of treatment-related toxicity on patients. However, a significant proportion of patients with OPC are currently treated with upfront chemoradiation without undergoing neck dissection, resulting in the absence of formal pathological lymph node evaluation, thus motivating the use of imaging as a surrogate marker of ENE. 
        
\begin{figure}[t!]
            \centering
            \includegraphics[width =0.75\textwidth]{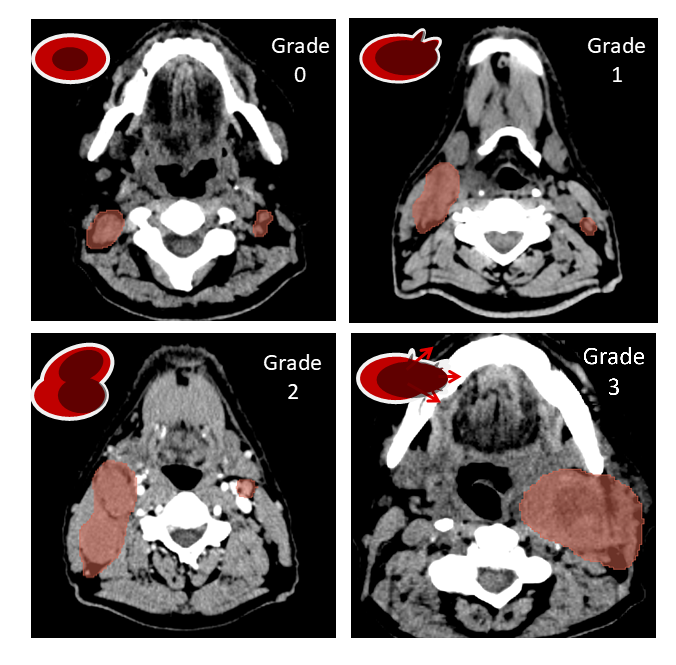}
            \caption{
                Illustration for iENE status grading system.
                Nodes are presented with white borders, red healthy tissue background and black tumoral content.
                Grade 0 indicates metastatic pathological node with no signs of extension.
                Grade 1 indicates extension through the nodal capsule into surrounding fat.
                Grade 2 characterizes coalescent nodal masses with loss of internodal planes.
                Grade 3 is defined by overt invasion of surrounding structures and tissues.
                In case of multiple ENE, the maximum grade is reported.}
            \label{fig:Introduction figure} 
        \end{figure}

        To expedite the evaluation process, imaging-based gradation systems for imaging detectable ENE (iENE) on head and neck CT scans have been developed in recent years, such as the five grade scale proposed by \cite{hiyama_extra-nodal_2020,OSulivanGradingSystem,OsulivanGrading2}. However, these lack standardization, and their clinical utility is hindered by inter-observer variability and inconsistent segmentation boundaries. Figure \ref{fig:Introduction figure} illustrates the four-tier iENE classification framework: grade 0— no radiologic evidence of ENE; grade 1— minor extension (through the capsule into surrounding fat); grade 2— coalescing matted nodal mass with disappearance of inter-nodal planes; and grade 3— gross invasion into surrounding anatomical structures (adapted from \cite{OSulivanGradingSystem,OsulivanGrading2}).  
        
        Therefore, despite increasing evidence advocating for the incorporation of ENE status into the staging of HPV-associated OPC, numerous challenges ranging from diagnostic inconsistency to a lack of universal criteria continue to impede its widespread and reliable integration into oncologic practice.
    
        In recent years, there has been growing interest in leveraging machine learning (ML) techniques to extract valuable prognostic insights from medical imaging, providing a way to extract predictive information beyond subjective image interpretation and potentially enhancing clinical guidelines with more precise and reliable image analysis. By leveraging the complementary predictive value of multiple modalities and their interactions, patient-specific treatment strategies could be integrated and result in reduced toxicities (\cite{boehm_harnessing_2022}).
        
        In this paper, we propose AMO-ENE, a fully automated pipeline using an attention-based multi-organ fusion model for for outcome prediction in head and neck oropharyngeal cancer, focusing on primary tumors and their association with metastatic extranodal extension (ENE) in CT images. This study proposes a novel prognostic pipeline that includes an ENE segmentation model and introduces a multimodal and multi-omic attention fusion model for multi-bin outcome risk prediction.  Our objective is to propose an end-to-end, clinically applicable pipeline, with efforts on exploring stronger deep modality fusion algorithms paired with deep survival methods. This paper’s contributions include the following:
        
        \begin{itemize}
            \item Automation of pathological node segmentation on radiation planning CTs, comparing state-of-the-art methods, and subsequent iENE grade classification based on complementary feature sets;
         
            \item Introduction of an attention-based multi-omics fusion model for 2-year risk assessment in HPV-associated OPC treated with radiation therapy;
            
            \item Adaptation of the proposed model to multi-bin risk modeling for a deep learning based-longitudinal risk estimation of outcomes in HPV-associated OPC;
            \item Evaluation of extra nodal extension prognostic value using statistical survival analysis on a cohort of 397 OPC patients.
        \end{itemize}
    
    \section{Related Works}
    
    \subsection{Lesion segmentation}
    
        Medical image segmentation is a popular task in oncological research, as robust tumor detection and  delineation may enable automated screening and treatment planning processes. In recent years, several approaches were proposed to automatically retrieve organs at risk (OARs) from planning images \cite{peng_oar-unet_2024}, however they often exclude nodal structures in this category. In addition, there are no methods to our knowledge for automated extra nodal extension segmentation in the literature, which may be explained by the lack of iENE status integration in clinical guidelines for HPV-associated OPC and by the labour intensive annotation task it demands.
        
        Since its introduction in 2012 \cite{AlexNet}, deep convolutional neural networks (CNN) have established themselves as a standard for computer vision tasks. Adapted to the popular encoder-decoder "U" style architecture \cite{Unet}, it varies to different field of views by harnessing the strong spatial awareness of convolution masks in an encoder-decoder fashion for 3D segmentation prediction. The 2022 HECKTOR MICCAI challenge \cite{Hecktor} demonstrated the value of  CNN models for nodal lesion segmentation, \cite{SegResnet} proposed a residual CNN for this task and achieved the best overall results. In addition, \cite{rebaud_simplicity_2022} provided a  nnUnet pipeline \cite{nnunetv2}  for gross tumor volume (GTV) and nodal lesion segmentation. This self - configuring U-shaped convolutional neural network, known for its reliability and state-of-the-art performances, allowed for automated parameter tuning and preprocessing, often serving as a strong segmentation baseline in comparative studies.
        
        Vision  Transformer (ViT) \cite{Vit+CNNHNC} 
        \cite{benchmarkVitHNC} segmentation architectures, with their ability to model long-range dependencies, have been proposed as an alternative to convolution networks, leveraging attention  between input image patches treated as tokens. SwinUNETR\cite{SwinUNETR} has been proposed as a U-shaped iteration of the ViT, integrating a shifted windowed attention mechanism allowing multi-scale refinement of this mechanism to capture finer details. SwinUNETRV2 \cite{SwinUNETRv2} is an incremental modification of the aforementioned network, with an hybrid CNN-Vit encoder, combining the strength of both mechanisms with reduced data requirements.  
        
        Datasets with  segmentation annotations from large cross institutional cohorts have recently led to the popularity of foundation models, which can leverage large scale pre-training datasets for efficient fine-tuning capability. The generalist segment anything model (SAM) \cite{SAM} was adapted to a medical imaging oriented 3D version in SamMed3D \cite{SamMed3D}, a prompt-based 3D Foundation model pre-trained on 143 thousand segmentation masks. Foundation models offer the ability to segment organs accurately via the given prompt \cite{SamMed} \cite{MA-SAM}, even for previously unseen structures. However SamMed3D requires location prompts as input, effectively skipping the detection task inherent to segmentation. 
    
    \subsection{Node and extra nodal extension grade classification}
    
        Several approaches have explored the characterization of lymph node lesions in OPC, albeit not considering iENE as a primary target. In addition to the lack of ENE segmentation models, there are still no imaging based extra nodal extension classification model in the literature.
        
        Kann et al. \cite{kann_pretreatment_2018} developed a three-dimensional convolutional neural network (3D-CNN) to detect pathologically confirmed extranodal extension on head and neck CT scans across various head and neck cancer (HNC) subtypes. While the work also focused on the automated ENE prediction, authors utilized pENE—determined via histopathological analysis—as the ground truth, whereas current efforts have explored the consensus of radiologists for imaging-detected ENE (iENE) as the target variable, as neck dissection are seldomly performed as the initial therapy. Although pENE offers a more definitive standard by capturing microscopic disease indicators not always apparent on imaging, its clinical relevance in HPV-positive oropharyngeal squamous cell carcinoma (OPSCC) is often debated, with evidence suggesting that microscopic pENE may not significantly influence outcomes \cite{ECS,ECS2}. In addition Kann et al. did not assess whether the proposed AI-predicted pENE correlated with clinical endpoints.

        The Hecktor 2022 challenge \cite{Hecktor} catalyzed further exploration into the integration of nodal radiomics and multi-omic data for enhanced outcome prediction. Notably, the top-performing team \cite{rebaud_simplicity_2022} achieved a concordance index (C-index) of 0.68 for progression-free survival prediction in HNC. The winning approach involved the extraction of 1,209 features from both primary and nodal tumor segmentation, subsequently analyzed through a bagged binary weighted model. Other contributions in the challenge investigated the fusion of segmentation data with radiomic descriptors \cite{meng_radiomics-enhanced_2022}; however, these methods did not surpass the predictive performance of radiomics alone. While these studies confirm the predictive benefit of combining features from primary and nodal lesions, they notably omit extra nodal extension, a clinically relevant factor in prognosis.
        
    \subsection{Survival analysis for OPC}
    
        Survival analysis is a statistical framework aimed at estimating risk scores associated with patient outcomes, primarily by forecasting time-to-event variables, such as disease progression or mortality.
        
        In the application to HNC, radiomic features have demonstrated significant prognostic value in survival prediction tasks \cite{vallieres_radiomics_2017}. These characteristics, extracted from medical images, have proven effective in characterizing the morphological and textural attributes of primary tumor volumes, thereby offering predictive insights into adverse treatment outcomes. Several methods have been proposed to harness the integration of radiomic data, baseline logistic regression models \cite{RadCure} with traditional lasso-based feature selection, non-parametric bagged estimators with a signed attribution to each feature \cite{rebaud_simplicity_2022} or other random forest algorithms with recursive feature selection \cite{vallieres_radiomics_2017}. More recent approaches have integrated radiomic features in deep networks \cite{saber_feature_2024}, alleviating the curse of dimensionality with strong regularization and higher non-linear parameter spaces. Other works proposed multimodal fusion across radiology, pathology, and genomics,  \cite{vanguri2022multimodal}, but using simple fusion strategies which may hinder generalization. 
        
        Emerging methodologies have begun to shift focus away from handcrafted radiomic features, exploring the latent potential of non-explicit representations. 
        
        Vision-based deep learning models have been explored, such as salient features extractors from images \cite{Presanet}, allowing for a deeper exploration of hierarchical biomarkers.
        In particular, foundation models using self-supervised learning (SSL) strategies \cite{fmcib} have been proposed. These models leverage vast amounts of unannotated data to learn robust, high-dimensional representations without the need for manual feature engineering. While they offer a promising avenue for generalizable and multi-task frameworks in biomarker discovery, their opaque nature as “black boxes” makes the interpretability of their outputs limited. This is of primary concern in the safe use of critical medical applications. Importantly, their application in integrating nodal characteristics, particularly in HPV-associated oropharyngeal carcinoma (OPC), remains unexplored. Recent work has used a radiomics-based method for ENE prognosis in OPC patients, but focused only on single source integration \cite{dayan2025artificial}.
        
        Modeling time-to-event outcomes is often performed using Cox proportional hazards regression, where hazard ratios derived from input features serve as predictors of event timing. While Cox regression remains a widely used and effective method for survival analysis, as it relies on the proportional hazards assumption and can be sensitive to high-dimensional or correlated features \cite{CoxPrognostic}, which may limit its applicability in complex clinical datasets.
        
        To address these limitations, Multi-Task Logistic Regression (MTLR) was introduced by \cite{MTLR_base} as a discrete-time survival modeling framework that enables flexible risk estimation without relying on the proportional hazards assumption. MTLR models the probability of surviving each time interval as a sequence of logistic regressions, thereby allowing to adapt to various heterogeneous data types. Its effectiveness has been demonstrated in clinical contexts, including head and neck cancer prognosis \cite{MtlrOPC}, where it outperformed traditional survival analysis methods.

\section{Methods}\label{Materials and Methods}
    
\begin{figure*}[th!]
            \centering
            \includegraphics[width =1.05\textwidth]{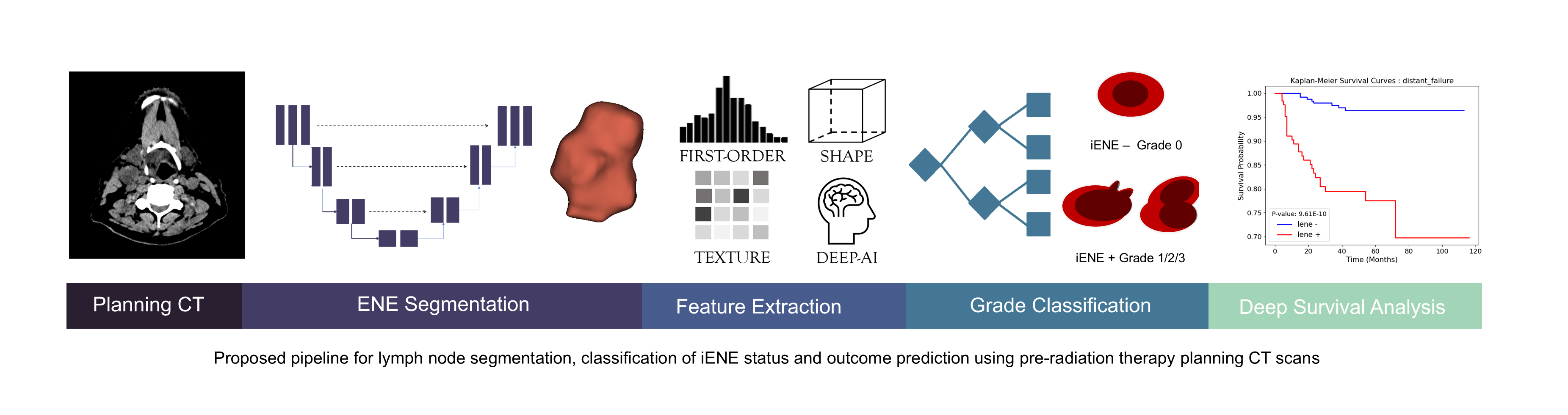}
            \caption{
            Overall schematic diagram of the proposed framework for CT-based iENE status and prognosis prediction.
            First, an automated segmentation and selection of pathological node allows for the retrieval of the largest metastatic lymph node.
            Afterwards, radiomics and deep features are extracted from the structure to classify iENE grades in a binary problem.
            Finally, from the obtained predictions, the survival characteristics of predicted iENE groups are compared to demonstrate the influence of extra nodal extensions in HPV-associated OPC.}
            \label{fig:Pipeline}
        \end{figure*}

        In this section, we describe the patient cohort characteristics, the overall prediction framework for iENE and its individual components.
        Figure \ref{fig:Pipeline} presents the overall CT to outcome pipeline.
        In Sec. \ref{Dataset collection}, we report  the detailed patient characteristics and 
        inclusion/exclusion criteria.
        Sec. \ref{Segmentation Method}  describes the iENE segmentation module implementation and training parameters, while in Sec. \ref{Grade Classification} we list the 
        different feature extraction methodologies and the subsequent ENE grade classification
        experiments, followed by the survival analysis methodology (Sec. \ref{Survival Methodology}) and the multi-organ attention based outcome prediction model.
        Finally,  Sec. \ref{Setup Methodology} reports the imaging and experimental setup  used in each step of this study.
    
    \subsection{Clinical datasets}\label{Dataset collection}
    
        Following institutional review board approval, we collected a retrospective cohort treated at Centre Hospitalier de l'Université de Montréal (CHUM) for HPV-associated OPC between 2009 and 2020. The study cohort comprises 397 patients diagnosed with oropharyngeal cancer, with a median age at diagnosis of 62 years (range: 39–84). The HPV status was clinically established with a positive p16 status upon biopsy of either the primary tumor or cervical lymph node. Exclusion criteria includes non curative intent chemo-radiation, negative HPV status, non cN+ staging and transoral robotic surgery (TORS) treated patients.
        
        A vast majority of patients were male (317 males vs. 80 females). Median number of smoking pack-years was 8 (range: 0–100), with 46 $\%$ current smokers and 67$\%$ reporting prior exposure to tobacco. Tumor staging, based on the AJCC 8th edition \cite{AJCC8TH}, indicates that 25.6 $\%$ of patients had T1 tumors, 38 $\%$ had T2, 24 $\%$ had T3, and 12.4 $\%$ had T4. Nodal staging showed a predominance of N1 disease (311 patients), followed by N2 (67 patients) and N3 (19 patients). Concurrent systemic therapy was administered to the majority of the cohort (333 patients).
        
        Most patients had between 1 to 4 abnormal lymph nodes (298 patients), while 99 patients presented with 5 or more. Abnormal retro-pharyngeal lymph nodes larger than 8 millimeters were observed in 44 cases. Radiological assessment of extra nodal extension (iENE) showed that 271, 26, 75, and 25 patients were scored as iENE 0, 1, 2, and 3 respectively. 
        
        For all patients, the following information was collected and represented the clinical variables incorporated in the prognosis prediction model: 
        \begin{itemize}
            \item Age at diagnosis: Patient age before treatment planning;
            \item Patient sex: Biological sex at diagnosis;
            \item ECOG \cite{ECOG} score: A standardized measure of cancer's impact on patient functional status;
            \item Smoking pack-years: Measures lifetime tobacco exposure;
            \item TNM 8th: Clinical AJCC staging \cite{AJCC8TH} for Tumoral, Nodal and Metastatic components;
            \item Concurrent Systemic Chemotherapy: If the patient undergoes chemo-radiation;
            \item Type of Concurrent chemotherapy: Protocol used in the chemotherapy.
        \end{itemize}
        For each patient, the planning pre-treatment CT scan and the associated radiation oncologist segmentations of gross tumor volume and iENE were collected.
        
        Therapeutical outcomes were collected at study date, leading to the report on the population observed events in the following Table \ref{tab: Outcome Repartition}.

        \begin{table}[h]
            \centering
            \begin{tabular}{@{}ccccc@{}}
                \toprule
                Outcome              & Event type & Male & Female & Total \\ \midrule
                \multirow{2}{*}{OS}  & Uncensored & 27   & 11     & 38    \\
                                     & Censored   & 290  & 69     & 359   \\ \midrule
                \multirow{2}{*}{DM}  & Uncensored & 30   & 5      & 35    \\
                                     & Censored   & 287  & 75     & 362   \\ \midrule
                \multirow{2}{*}{DFS} & Uncensored & 52   & 16     & 68    \\
                                     & Censored   & 265  & 64     & 329   \\ \bottomrule
            \end{tabular}
            \caption{
                Outcome representation in the patient cohort for Uncensored (Event presence at the latest available follow up) and Censored patients, stratified for different outcome measures 9OS, DM, DFS).}
            \label{tab: Outcome Repartition}
        \end{table}
    
        Here, overall survival (OS) was defined by mortality from any cause, distant metastasis (DM) by the failure at distant sites from the head and neck region and disease-free survival (DFS) by any recurrence or death at follow up. Time to event was calculated from date of diagnosis. The mean and median follow-up were 47.0 (±22.3) months and 44.4 (inter-quartile range 32.7-61.1).
        
        \paragraph{\textbf{iENE Annotation Protocol}}
        
        Ground truth imaging‐detected extra nodal extension (iENE) was graded and segmented independently and blinded to clinical outcomes by two board‐certified neuroradiologist. In addition, fifty five head and neck CT cases were independently assessed by three raters to rate agreement on decisions.
        
        The distribution of graded iENE presents as a heavily unbalanced problem, with  69.4\%, 5.9\%, 18.8\% and 5.9\%, proportion for grades 0 to 3 respectively.
    
    \subsection{Segmentation of primary of ENE nodules}\label{Segmentation Method}
    
    \subsubsection{Hierarchical SwinUNetR segmentation of iENE}
    
        To automatically segment the ENE's on CT scans, we proposed a model inspired by the SwinUNETRV2 framework \cite{SwinUNETRv2}, which incorporates a collection of U-shaped CNN networks with automated processing and augmentation transformations selection, combined with a Swin Transformer backbone. We selected the 3D residual encoder model (M version) for the CNN components, and trained it on the entire head and neck region of interest (ROI), processed in a sliding patch window process, in order to segment all pathological node extensions. To address challenges in scarcity of high-quality annotated data and significant variability across cases, the model includes a SSL workflow that integrates a hierarchical vision transformer with a multi-scale masked auto-encoder (MAE) pretraining strategy for enhanced representation learning as shown in \cite{he_masked_2021}.
        
        By leveraging 3D convolutions at varying depths, the encoder-decoder CNN architectures allows to integrate local and global features, enabling precise object detection and border delineation. This process is particularly  relevant for nodal extensions, since  precise local decisions are required to accurately evaluate the spread of tumor content.
        
        We trained the binary semantic segmentation model using the
        soft Dice Score (DSC) loss as described in 
        Eq. \ref{eq:DiceScoreEquation}:
    
         \begin{equation}
            \mathcal{L}_{\text{Dice}} = 1 - \frac{2 \sum_{i=1}^{I} Y_i \hat{Y}_i}{\sum_{i=1}^{I} Y_i + \sum_{i=1}^{I} \hat{Y}_i + \epsilon}
            \label{eq:DiceScoreEquation}
         \end{equation}
    
        where $Y_i$ represents the ground truth, \( \hat{Y_i} \) 
        the predicted voxels for each patient $i \in I$, and $\epsilon$ ($1\times10^{-8}$) a stability term.
    
    \subsubsection{Nodal component selection}\label{NodalSelection}
    
        Due to significant segmentation margin variability between multiple nodal components in the head and neck region with the ground-truth data, simply selecting the largest predicted segmentation mask as the main nodal component may lead to incorrect identifications of pathological nodes. This is largely prominent in the case of lower grade statuses, as the effect of ground-truth segmentation margins is more likely to cause matching errors in lower volume nodes.
         
        To address this issue, we propose an iENE selection method based on a node comparison algorithm and uncertainty estimates, aiming to match the actual largest ground truth with its predicted component. To achieve this, the algorithm  analyzes all the node volumes from the $N$ segmented components. Given a predicted a series of $N$ segmented components, the algorithm selects the largest component, only if the percentage difference between the top volume and next-largest detected ganglion pairs falls below a threshold $\rho$. Here, $\rho$ represents the volume difference between pairs nodes. In case of multiple nodes fall below $\rho$, the algorithm selects all nodes falling within the $\pm$2SD range of segmentation variability. From the selected sub-set, the node with  the lowest uncertainty measure of all possible ganglions intersecting with the predicted nodule is chosen. The segmentation uncertainty approach is based on Sparse Baysian Models \cite{abboud2024sparse}, which selectively assigns a subset of parameters as Bayesian by assessing their deterministic saliency through gradient sensitivity analysis, providing a quantitative measure of uncertainty from ground-truth annotations.

        \begin{figure*}[th!]
            \centering
            \includegraphics[width = \textwidth]{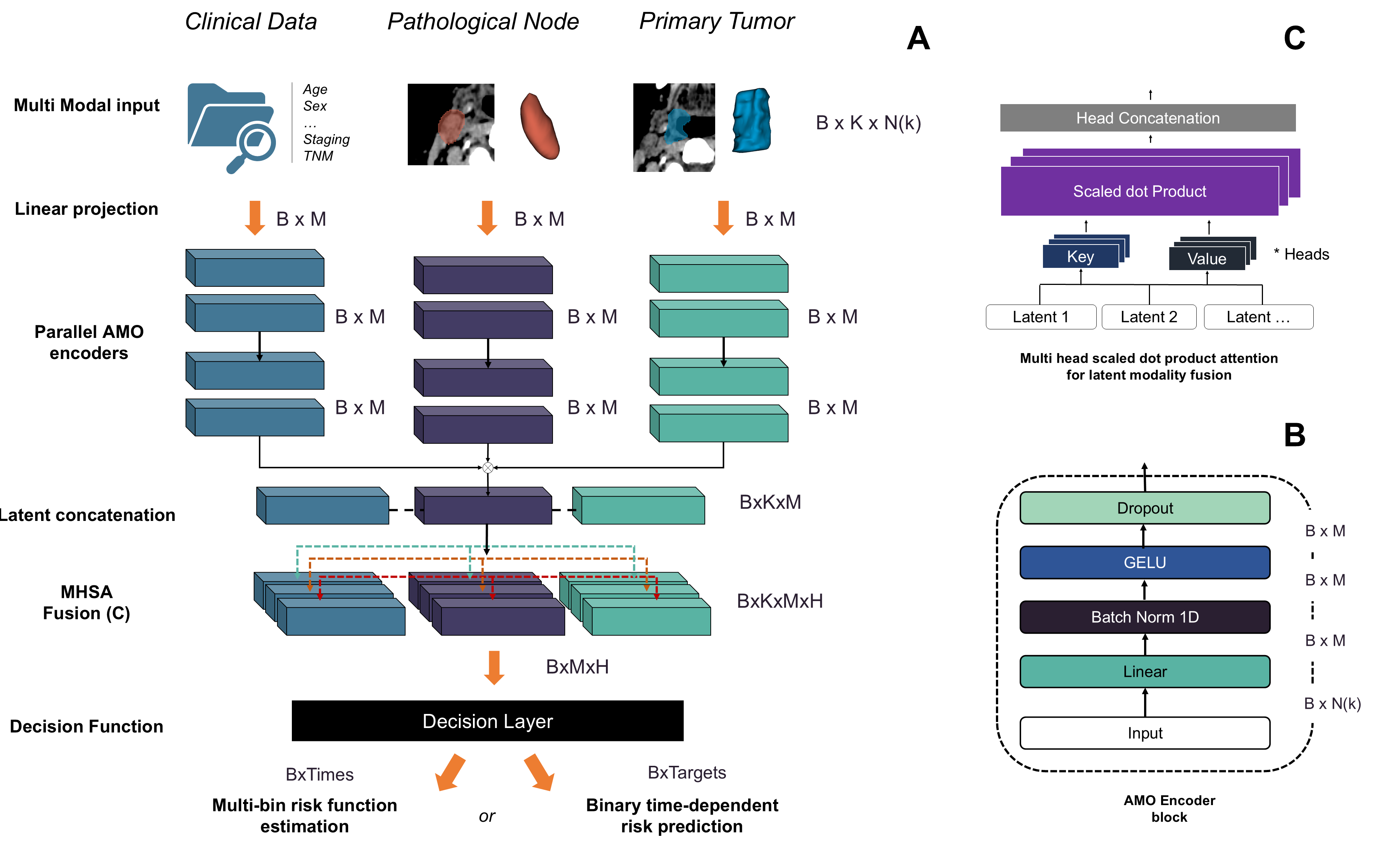}
            \caption{
                Graphical representation of the proposed attention-based modality fusion pipeline.
                In (A), multiple modality-specific feature encoders are trained in parallel, to extract features from each specific modality.
                The latent representation of each modality is concatenated and a multi-head scaled dot product attention is performed to evaluate interactions and combine latent representations.
                A final decision layer is adaptable to either predict a time dependent outcome target or multi-bin risk modeling.
                Details are provided in Sec. \ref{Subsection BOM} for relevant notations.
                In (B), the proposed modality encoder module is shown, which is composed of multiple blocks containing  parametric layers.
                (C) Detailed description of the operation performed in the attention layer. }
            \label{Figure:Model}
        \end{figure*}

    \subsection{Feature extraction and classification}\label{Grade Classification}
    
        From the generated segmentation masks for extra nodal extension, we aim to extract specific features allowing the automatic and robust identification of the extension status in patients.
    
        \paragraph{\textbf{Dichotomization schemes}}
    
        ENE grading presents itself as a heavily unbalanced problem with up to 10 times more class 0 than the other three and edge cases presenting mostly between classes 0-1 and 2-3 further increase potential variability in the dataset. These edge cases result mainly from ambiguous findings. As such, we propose three distinct dichotomization schemes, as opposed to a multi-class problem.
        
        We refer to the 0 vs 1-2-3 scheme as iENE$^-$ vs iENE$^+$, and it is the basis to all further model development with outcomes. This case maximizes the retention of ENE nodes, regardless of their severity. We also evaluate the 0-1 vs 2-3 classification scheme, which may represent a more conservative gradation of ENE, especially relevant in potentially indecisive radiologist annotations. Finally, we also classify grades 3 against other grades, placing greater emphasis towards detecting aggressive features and nodal extensions.
        
        \paragraph{\textbf{Feature extraction}}
        
        From the segmented ENE node and GTV masks (obtained from planning), we propose two distinct feature extraction approaches based on hand-engineered characteristics and deep self-supervised features to classify and anticipate outcomes. 
        For the hand-engineered approach, we extract radiomic features from the CT with the Python PyRadiomics \cite{pyradiomics} package. We set intensity bins to 10HU and resample instances to a fixed isotropic $1\times1\times1$ mm spacing. Computed features are obtained by selecting the first-order, shape, and gray-level co-occurence matrix (GLRLM, GLSZM, GLDM and GLCM) features (sum average disabled). A total of a 100 features were computed describing the shape of the predicted node, the intensity statistics, and high level texture maps.
        For the self-supervised approach, we extract deep features obtained from a pre-trained foundation model \cite{fmcib} for cancer imaging biomarkers (referred to as FMCIB). This method extracts 4096 features in the latent space of the self-supervised model which was pre-trained on 11 000 lesions. A fixed region of interest of size $50\times50\times50$ mm is extracted around the center of mass of the predicted iENE segmentation and fed to the model. Prior to inference, all scans are resampled to isotropic $1\times1\times1$ mm spacing and normalized by subtracting -1024 and dividing by 3072 (lower and upper bounds provided by the FMCIB model \cite{fmcib}).
        
        \paragraph{\textbf{Grade classification}}
        
        Using the extracted features, we construct a modular binary classification pipeline with components that were systematically evaluated through a grid search (see Sec. \ref{Setup Methodology} for all evaluated hyperparameters).
        
        All tabular data underwent standard scaling based on the respective training set distributions. We trained machine learning classifiers such as the Random Forest (RF), the XGBoost technique \cite{chen_xgboost_2016} and a deep learning based multi-layer perceptron (MLP).
        
        For each method, we tested out principal component analysis (PCA) and Lasso regression for feature selection. This step is crucial as both radiomic and FMCIB features present high co-linearity, increasing noise during model training. To handle class imbalance, we employ SmoteTomek \cite{Smote} generation of training samples, which allows to balance both classes while preserving variance in the under-represented class. Grid search parameter evaluation was performed for each combination of classifier and selection algorithm, where parameters from both components in each cross-validation fold were optimized to prevent data leakage.

    \subsection{Attention-based Multi-modal Outcome (AMO) risk prediction model}\label{Subsection BOM}
    
        In order to combine our multi-modal radiomic feature sets, we propose AMO-ENE, a scalable multi-branch attention-based deep classifier module, allowing for the independent unimodal representation extraction of biomarkers paired with the fusion of these modalities in order to assess interactions and their relative importance with respect to prognostic prediction.
        
        Furthermore, to explore and integrate all available modalities, namely clinical features and nodal radiomic/deep features extracted (detailed in Sec. \ref{Grade Classification}), we integrate primary tumor characteristics using the same radiomic and foundation model pipeline. This evaluation is motivated by the hypothesis that nodal features are complementary with primary tumor characteristics in their discriminative ability to assess the metastatic potential of HPV positive OPC.
        
        Let $\mathcal{X} = \{X^{(1)}, X^{(2)}, \dots, X^{(K)}\}$ denote a set of $K$ input modalities, such as radiomic features extracted from multiple organs or lesions. Each modality $X^{(k)} \in \mathbb{R}^{n_k}$ represents a feature vector of dimensionality $n_k$. The multimodal representation is derived via the following steps:
        
        \vspace{10pt}
        \noindent \textbf{1. Image-feature extraction} Each modality is processed by a modality-specific neural encoder $f^{(k)}$ to project it into a shared latent space of dimension $M$:
        
        \begin{equation}
            C^{(k)} = f^{(k)}(X^{(k)}) \in \mathbb{R}^{M}, \quad \forall k \in \{1, \dots, K\}.
        \end{equation}
        
        \noindent Each encoder $f^{(k)}$ is implemented as a two-layer feedforward neural network. The architecture consists of a first linear transformation mapping from $\mathbb{R}^{n_k}$ to $\mathbb{R}^{M}$, followed by batch normalization, a GELU (Gaussian Error Linear Unit) activation, and dropout for regularization. This consistent structure ensures each modality is independently projected into a common latent space while mitigating overfitting and capturing non-linear relationships. These encoders serve the dual role of feature selection and dimensionality reduction, yielding modality-specific latent embeddings $C^{(k)}$.
        
        \vspace{10pt}
        
        \noindent \textbf{2. Latent concatenation} The latent embeddings from all $K$ modalities are then vertically stacked to form a matrix $C \in \mathbb{R}^{K \times M}$:
        
        \begin{equation}
            C = \begin{bmatrix}
            (C^{(1)})^T \\
            (C^{(2)})^T \\
            \vdots \\
            (C^{(K)})^T
            \end{bmatrix}.
        \end{equation}
        
        \noindent Here, each row of the matrix $C$ corresponds to the latent embedding of a specific modality, where $C^{(k)} \in \mathbb{R}^M$ is transposed to align with row-major stacking. The resulting matrix $C$ thus encapsulates modality-wise biomarker representations, where each modality contributes one row of $M$ latent features.
        
        \vspace{10pt}
        \noindent \textbf{3. Multi-Head Self-Attention fusion (MHSA)}  
        To capture dependencies between modalities and enhance contextual interactions, we employ a self-attention fusion mechanism based on multi-head attention, followed by residual connections and layer normalization.
        
        Given the stacked modality representations $C \in \mathbb{R}^{K \times M}$ where each of the $K$ rows corresponds to a latent embedding of a modality in the $M$ dimensional latent space the self-attention layer operates as follows. The input $C$ is treated as the set of queries, keys, and values:
        
        \begin{equation}
            \hat{C}, \; A = \text{MultiHeadAttention}(Q=C, K=C, V=C),
        \end{equation}
        
        \noindent where $A \in \mathbb{R}^{K \times K}$ denotes the attention weight matrix capturing pairwise dependencies among modalities, and $\hat{C}$ the attention output of shape $\mathbb{R}^{K \times M}$.
        
        A residual connection is applied between the attention output and the original modality embeddings, followed by layer normalization to stabilize training. The resulting representation is further refined using a two-layer feedforward network with GELU activation and dropout.
        
        \vspace{10pt}
        \noindent \textbf{4. Global fusion for downstream tasks}  
        Finally, following attention-based interaction and refinement, we aggregate the modality-enhanced embeddings to produce a single unified representation via average pooling across modalities:
        
        \begin{equation}
            \bar{Z} = \frac{1}{K} \sum_{k=1}^{K} \hat{C}_k \in \mathbb{R}^{M},
        \end{equation}
        
        \noindent where $\hat{C}_k$ is the $k$-th row of the final attention-refined matrix $C$. The resulting vector $\bar{Z}$ captures the integrated multimodal information and serves as input to downstream predictive modules (e.g., classification or regression heads).

        Figure \ref{Figure:Model} presents an illustration of the proposed module. Each feature-set is given as input to its dedicated encoding branch, composed of two successive extractor blocks that combine high level features to a task specific biomarker latent. Each modal representation is concatenated and fused through a multi-head scaled dot product attention layer, allowing the retrieval of task specific interactions and serving as an attention weighted feature selection mechanism.
        
        The decision layer of the model can be adapted to multiple prognosis prediction paradigms, we thus propose two distinct cases in the following paragraphs describing binary time assessment, and multi-bin risk modeling.

        \paragraph{\textbf{Multi-modal 2-year risk/prognosis prediction}}
        \label{Survival Methodology}
        
        The model was trained to predict the 2-year landmark, where time to outcome was determined from the treatment start date.
        
        Using as input to the three distinct branches  the nodal extension radiomics, the GTV radiomics and the clinical data, we predicted the presence (1), absence or censoring (0) of OS, DM, and DFS as binary oncological outcomes. 
        
        We optimize the model with a weighted binary right censored cross entropy loss as described in Eq. \ref{eq:WeightedBCE}: 
        
        \begin{equation}
            L_{c c}=\sum_i\left[\begin{array}{l}
            \mathbbm{1}_{T} \cdot w_0 \cdot \ln \left(f\left(X_i\right)\right) \\
            +\left(1- \mathbbm{1}_{T}  \right) \cdot  w_1\cdot  \ln \left( f\left(X_i\right)\right)
            \end{array}\right],
        \label{eq:WeightedBCE}
        \end{equation}
        
        where the indicator function $\mathbbm{1}_{T}$ represents outcome presence at time $T$ (two year in our case). Weighting parameters $w_0$ are computed by obtaining the inverse population ratios relative to each class on the relative training set of each instance.  
        
        \paragraph{\textbf{Multi-bin risk modeling for survival analysis}}
        
        Beyond binary classification, we extend our multi-modal model to perform survival analysis through a discretized, multi-bin risk regression framework. This is achieved by replacing the classifier head with a Multi-Task Logistic Regression (MTLR) module~\cite{MTLR_base}. Patient survival times are discretized into \(T\) quantile-based intervals derived from the empirical survival distribution of the training cohort. The model then outputs a sequence of time-specific logits, each representing the log-risk score for a given time interval.
        
        Training is performed by minimizing the MTLR negative log-likelihood loss, defined as:
        
        \begin{equation}
        \begin{aligned}
            \mathcal{L}_{\mathrm{MTLR}} = 
            &- \sum_{i \in \mathcal{I}_c} \log \left( \sum_{t: y_{it} = 1} e^{z_{it}} \right) \\
            &- \sum_{i \in \mathcal{I}_u} \sum_{t=1}^T y_{it} z_{it} \\
            &+ \sum_{i=1}^N \log \left( \sum_{t=1}^T e^{z_{it}} \right),
        \end{aligned}
        \label{eq:NegLogMtlr}
        \end{equation}
        
        \noindent where \( \mathbf{Z} \in \mathbb{R}^{N \times T} \) are the model’s predicted logits over \(T\) time bins for \(N\) patients, \( \mathbf{Y} \in \{0,1\}^{N \times T} \) are the encoded survival targets, and \( \mathcal{I}_c \), \( \mathcal{I}_u \) represent the sets of censored and uncensored samples, respectively. This formulation enables the model to learn a smooth, time-dependent hazard function, capturing the changing risk of event occurrence over time. During inference, the predicted logits can be converted into cumulative survival probabilities, enabling individualized survival curve estimation.
        
    \subsection{Experimental setup and implementation}\label{Setup Methodology}
    
        All models and experiments were carried out on a single Nvidia A6000 GPU with 48GB of available VRAM. Segmentation experiments were performed on a 12GB RTX Titan card with batch size scaling. Experiments were carried out in pytorch version 2.1.0 and python 3.11.5. Size occupied on disk by the dataset of 397 cases was 20 GB.  The source code of the proposed model is made publicly available at \underline {https://github.com/...}. 
    
    \subsubsection{iENE segmentation}
        We initially trained the proposed segmentation model for 750 training epochs, which allowed for convergence of the model using the internal multimodal our dataset. We trained the model with a stochastic gradient descent optimizer with an initial learning rate of 0.01, $1\times10^{-5}$ weight decay factor and the polynomial scheduler. Given the characteristics of the dataset, namely the varying number of nodes to segment and the size discrepancy, we adapted the performance evaluation in our segmentation task to account for possible bias towards larger nodes.

        The largest node is the main region of interest in the imaging grading process and the most relevant for prognosis evaluation, and as such, we report segmentation results based on the largest predicted and annotated nodes, based on the selection algorithm presented in Sec. \ref{NodalSelection}.

        Results were evaluated using the mean Dice score obtained over 5 fold cross validation splits. Given $\textit{Y}$ the ground truth and $\textit{$\hat{Y}$}$  the predicted voxels, we extract the largest components from each masks and compute their Dice as:
        
        \begin{equation}
            \text{Dice} = \frac{2 \times \text{TP}}{2 \times \text{TP} + \text{FP} + \text{FN}}
        \end{equation}
        
        where TP, FP and FN are the true positive, false positive and false negative number of voxels.
        
        Segmentation models were implemented within the Monai \cite{Monai} framework, with Pytorch-lightning for acceleration \cite{Falcon_PyTorch_Lightning_2019}.

    \subsubsection{iENE classification}

        The classifier for nodal extension grades were evaluated in a 5-fold stratified cross-validation scheme, where each split obtains similar grade distributions as the overall population. We evaluated the different models with the mean area under the receiver operating characteristic curve (AUROC) metric computed over the test folds, as it allows for a fine logit threshold selection, maximizing the respective recall and specificity contribution of each model.
        
        Grid search parameter evaluation was performed for each combination of classifier and selection algorithm. Searched parameters included: number of estimators $\in \{50, 100, 200, 500, 1000\}$; tree depth $\in \{3, 5, 7\}$ ; initial learning rates $\in \{0.01, 0.1, 0.2\}$; L1 ratio $\in \{0.001, 0.01, 0.1, 1\}$; number of PCA components $\in \{5, 10, 15, ..., 100\}$.

    \subsubsection{Clinical value of iENE as a prognostic biomarker}

        To evaluate the iENE's predictive performance for outcome prediction, we conduct the following experiments. 
        
        We first analyze the predicted iENE grade as a stand-alone clinical biomarker for outcome prediction. Stratification of the study population and comparison with current AJCC reported criteria is reported. We then report a 2-year prediction experiment for each of the OS, DM and DFS binary outcome, introducing a novel attention-based multi-modal fusion predictor as the backbone. Interactions between iENE and primary tumor volumes are evaluated through this Deep Learning based model. Finally we explore the use of this model on a longer follow-up period with multi-bin risk modeling.
        
        \paragraph{\textbf{Kaplan-Meier iENE score univariate estimation}}
        \label{Subsection KM}
        As  iENE presence is not currently used in the current AJCC recommendations for HPV positive OPCs, we aim to demonstrate that iENE presence is  associated with worsened outcomes. From the previously obtained grade classification, we preserve the testing logits from the ENE- vs ENE+ dichotomization scheme as a specific feature. This pipeline based on synthesized deep feature descriptor is compared in uni- and multi-variate models for outcome prediction in OPCs. First as a stand-alone variate, we fit a Kaplan-Meier estimator to stratify outcomes based on the predicted grade probability. Additionally, we perform a log-rank test between the predicted score and outcome presence to determine the outcomes associated with the iENE features.
        
        \paragraph{\textbf{iENE vs ENE inter-rater variability}}
        
        As the proposed iENE classification model was trained on established neuro-radiology guidelines, we compared the obtained prognostic value of the model predictions with individual raters annotation in order to assess the clinical performance.
        
        We compared the obtained prognostic value of our model predictions with individual raters annotations on a separate test set of 55 patients, unseen at training. These were independently scored by three radiologists via a Log-rank test for the outcomes DM, DFS and OS, using the annotation as an univariate predictor. All scores used are dichotomized in classes iENE$^-$ vs iENE$^+$.
        
        Additionally, we report on a simulated reader by selecting a random annotation from the set of three radiologists for each patient, then bootstrapping this operation over 10 000 sets, effectively estimating the performance of a trained neurologist by providing an average of the annotation styles. This experiment allows us to estimate a standardized error rate across annotating practices, as well as to assess how a single reader’s iENE grading would correlate with outcomes.
        
        \paragraph{\textbf{Two year risk prediction}}
        
        All outcome prediction models were evaluated in a 5-fold stratified cross-validation scheme, penalized by the negative binary cross entropy log loss function defined as follows: 
        
        \begin{equation}
            L_{\log}(y, p) = -(y \log (p) + (1 - y) \log (1 - p))
            \label{logloss classic}
        \end{equation}
        
        where $y$ represents the ground truth label and $p$ the estimated predicted probability. Models were trained and evaluated using the mean AUC metric.
        
        Baseline model parameters were as follows: two extractor blocks, 500 feature latent space size, two attention heads with hidden size 256, ReLU non linearity activation function, dropout ratio of 0.3, Adam optimizer with a batch size of 32 and starting learning rate of $1\times10^{-3}$.
        
        To determine the optimal parameter selections, we performed a grid search on the following parameter ranges of values: learning rate $\in \{0.01, 0.001, 0.007\}$; batch size $\in \{8, 16, 32\}$; hidden size of latent space $\in \{256, 512, 1024\}$; dropout ratio $\in \{0, 0.3, 0.4\}$; number of attention heads $\in \{1, 2, 3\}$.
        
        \paragraph{\textbf{Multi-bin risk outcome modeling}}
        \label{MTLR experiment details}
        
        Finally, the multi-bin risk prediction performance was assessed using the C-index obtained over the same 5 folds as the previous experiment on binary prediction. 
        
        Bins were computed as described in the MTLR paper\cite{MTLR_base}, yielding $\sqrt{|observations|}$ bins corresponding to the quantiles of the survival time distribution.
        
        The C-index provides an assessment of the ranking ability of their survival times based on predicted risks  and is calculated as follows:
        
        \begin{equation}
            Cindex=\frac{\sum_{i \neq j} 1\left\{r_i< r_j\right\} 1\left\{T_i>T_j\right\}\mathbbm{1}_{j}}{\sum_{i \neq j} 1\left\{T_i>T_j\right\}  \mathbbm{1}_{j}} .
        \end{equation}
        
        with $r_i$ the predicted risk score for a patient $i$ and $T_i$ their time to event or censoring, effectively reporting the proportion of concordant pairs of risk ranking between two distinct patients over all possible pairings.
        
        We performed the same grid search as described in the binary risk assessment, carried out on the Pytorch Lightning \cite{Falcon_PyTorch_Lightning_2019} framework, using the lifelines \cite{Lifelines} package in order to compute the C-index.
        
    \section{Results}

    \subsection{iENE segmentation}
    
        Table \ref{tab:results} presents the results from the iENE segmentation module using the SwinUNETR-based framework with MAE pre-training. Overall, the proposed SwinUNETR-derived model with added modules achieves the best mean performances across all ENE grades with a DSC of 78.4 $\pm$ 7.5, which is a significant improvement to nnUnet \cite{nnunetv2} (74.4 $\pm$ 8.9), SwinUNETRV2 \cite{SwinUNETRv2} (70.1 $\pm$ 7.7) and basic SwinUNETR \cite{SwinUNETR} (69.5 $\pm$ 8.3). All fine-tuned models present a high variability in the performance of the segmentation, with standard deviations above 25 $\%$ for all of them.

        \begin{figure*}[tbh]
            \centering
            \includegraphics[width=1\textwidth]{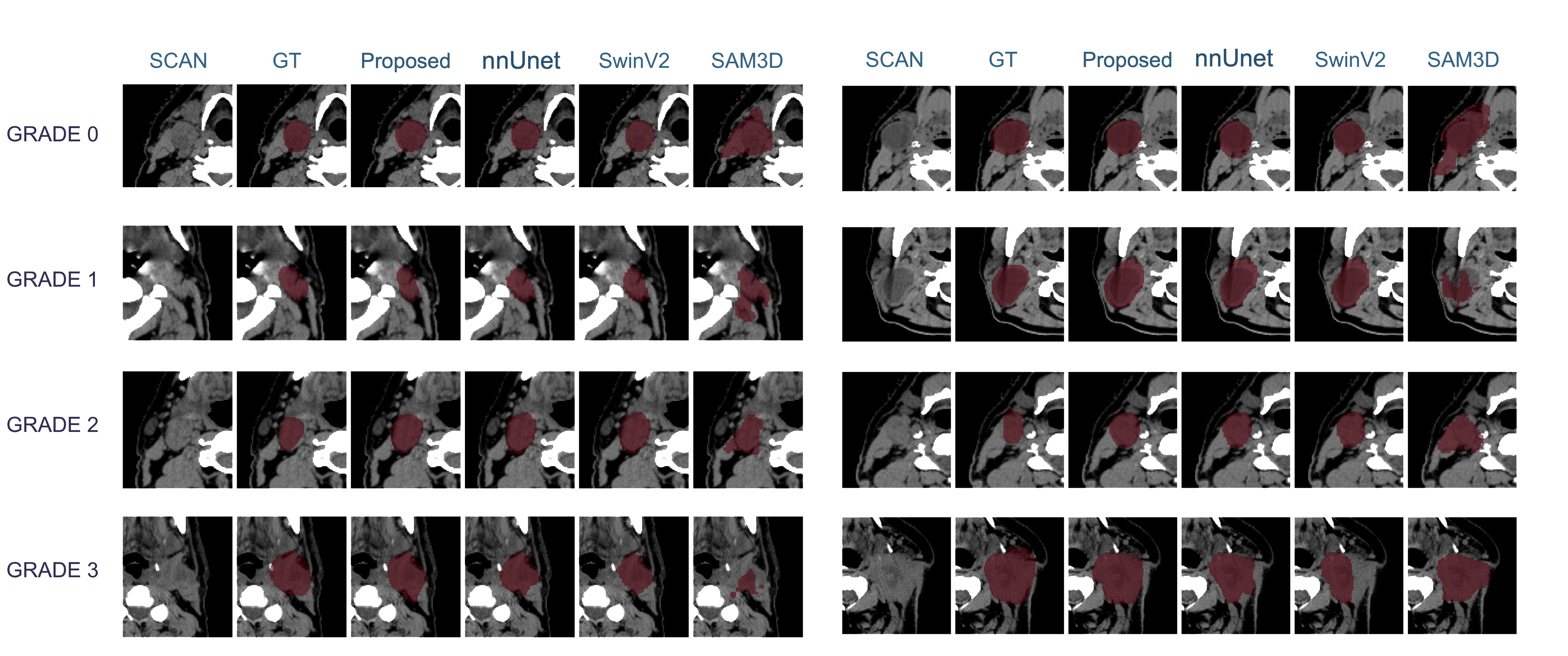}
            \caption{
                Qualitative results for the iENE segmentation task on radiotherapy planning CT images.
                Each row presents the raw CT scan and the output segmentation maps for two distinct patients from test folds over a 5-fold cross validation.
                Comparative methods are shown in their respective columns : proposed model,  nnUnet \cite{nnunetv2} ,  SwinUNETRV2 \cite{SwinUNETRv2}, and SamMed3D \cite{SamMed3D}. GT: ground truth mask.}
            \label{Figure:SegResults}
        \end{figure*}
        
        ENE grade stratification shows that smaller nodes (grades 0 and 1) have the  highest standard deviations ($9.0$ and $8.2$). Grade 2 nodes, on the other hand, show the best segmentation performance across all nodes and trained models, with a mean DSC of 83.4 $\pm$ 6.4 achieved by the proposed model. Grade 3 nodal extensions generally present worse segmentation results with high variance (70.2 $\pm$ 4.5 for the proposed pipeline), demonstrating the added difficulty in identifying the exact nodal extension margin on lower contrast regions where tissue and tumor content are of similar densities (refer to the last row of Figure \ref{Figure:SegResults} for a quantitative result).
        
        SamMed3D \cite{SamMed3D} performed the worst out of all models, with a Dice score of 46.4 $\pm$ 12.4 for the single prompt version. Increasing the number of prompt points to 10 did not improve performance, which lowered the mean Dice score to 39.4 $\pm$ 11.1.
        
        Qualitative results for the trained models are shown in Figure \ref{Figure:SegResults}. We can observe that the proposed segmentation model yields more accurate results, with detailed margins in comparison to other vision transformers, which can be explained by differences in receptive fields, with the local precision of convolution in comparison to the global context of the attention mechanism. For all these cases, the obtained segmentation detected the pathological nodule and provided out of object margins, as expected from annotations originating from radiotherapy planning. For the presented grade 3 case, we can observe that the proposed extremities of the extension vary significantly from the ground truth to models, showing the annotation difficulty of this advanced stage.
        
        \begin{table*}[h]
            \centering
            \begin{adjustbox}{max width=\textwidth}
            \renewcommand{\arraystretch}{1.3}
            \setlength{\tabcolsep}{4pt} 
            \begin{tabular}{ccccccccc}
            \hline
            Model &
              Type &
              Pre-T &
              Params &
              Grade 0 &
              Grade 1 &
              Grade 2 &
              Grade 3 &
              Average \\ \hline
              {nnUnet} \cite{nnunetv2}  &
             CNN &
               &
              101.9 M &
              68.8 ± 9.0 &
              68.1 ± 8.2 &
              80.3 ± 6.2 &
              68.0 ± 8.5 &
              74.4 ± 8.9 \\  
            SwinUNETR \cite{SwinUNETR} &
              Vit &
              $\checkmark$ &
              62.2 M &
              67.3 ± 9.9 &
              63.1 ± 10.0 &
              80.0 ± 7.6 &
              67.3 ± 7.9 &
              69.5 ± 8.3 \\
            SwinUNETRV2 \cite{SwinUNETRv2} &
              Vit + CNN & 
               &
              72.8 M &
              67.8 ± 10.0 &
              {69.3 ± 7.9} &
              {81.1 ± 8.4} &
              66.1 ± 6.1 &
              70.1 ± 7.7 \\
            SamMed3D (1pt) \cite{SamMed3D}*   & Vit + Prompt & $\checkmark$ & 100.0 M & 43.4 ± 12.6 & 55.1 ± 13.4 & 55.1 ± 10.7 & 58.3 ± 13.8 & 46.4 ± 12.4 \\
            SamMed3D (10pts) \cite{SamMed3D}*  & Vit + Prompt & $\checkmark$ & 100.0 M & 35.5 ± 13.0 & 37.9 ± 15.4 & 47.8 ± 13.6 & 54.9 ± 9.0 & 39.4 ± 11.1 \\
            \textit{Proposed pipeline} &
              Vit + CNN & $\checkmark$
               &
              79.4 M &
              \textbf{73.6 ± 9.0} &
              \textbf{71.3 ± 8.2} &
              \textbf{83.4 ± 6.4} &
              \textbf{70.2 ± 4.5} &
              \textbf{78.4 ± 7.5} \\ \hline
            \end{tabular}%
            \end{adjustbox}
            \caption{
                Quantitative segmentation results obtained in the largest iENE component segmentation task (n=397).
                Each result presents the mean DSC ($\pm$ standard deviation) over the 5 cross-validation folds.
                * SamMed3D is not trained and only used for inference with a selected number of interest point prompts; all other models are fine-tuned.
                \textit{Vit}: Vision image transformer architecture \cite{VIT}.
                \textit{CNN}: convolutional neural network architecture \cite{AlexNet}.
                \textit{Pre-T}: Number of overall neural network parameters.
                \textit{Pre-T}: Pretraining.}
            \label{tab:results}
        \end{table*}
    
    \subsection{Nodal component selection}
    
        In Table \ref{tab:ComponentSelResults}, we assess the impact of the component selection method and variants on the predicted pathological nodes in correspondence with their  ground-truth structures. As previously noted, the naive approach results in numerous cases where no component is detected, yielding a DSC of $78.4 \pm 7.5$. This issue is particularly prevalent in lower-grade iENE cases, where the relatively small nodal volumes make segmentation more susceptible to margin-based mismatches.
        
        \begin{table}[h]
            \centering
            \begin{adjustbox}{width=0.7\textwidth}
            \begin{tabular}{lcccc}
                \toprule
                Method & Precision & Recall & IoU & DSC \\
                \midrule
                Naïve                    & $78.4 \pm 7.0$ & $79.2 \pm 6.4$ & $67.9 \pm 7.9$ & $78.4 \pm 7.5$ \\
                $\rho = 40\%$               & $80.2 \pm 5.8$ & $81.5 \pm 5.8$ & $68.1 \pm 7.5$ & $80.8 \pm 6.7$ \\
                Best Match               & $82.2 \pm 4.8$ & $84.4 \pm 4.9$ & $72.1 \pm 6.1$ & $83.5 \pm 4.1$ \\
                \bottomrule
            \end{tabular}
            \end{adjustbox}
            \caption{
                Comparison of component selection methods for pathological node segmentation using the proposed SwinUNETRV2-based model.
                Metrics are reported as the mean and standard deviation across all test samples from 5-fold cross-validation.
                The naive method pairs the largest predicted and ground truth nodes.
                The IoU-based method computes DSC across all node pairs with a threshold set at $\rho = 40\%$.
                The best match method returns the highest DSC score among pairs that include the largest predicted node.}
            \label{tab:ComponentSelResults}
        \end{table}
        
        Adjusting the component selection threshold narrows the distribution of overall DSC values, moving its mean value from 78.4 to 83.5, indicating that some previously undetected cases were misclassified due to incorrect association. Overall, the model reaches a recall of 84.4 $\%$ ($\pm$ 4.9)  and a precision of 82.2 $\%$ ($\pm$ 4.8) of ENE voxels from masks associated with the largest predicted node.
    
    \subsection{Automatic iENE grading from segmented CT lesions}
    
\begin{figure*}[h]
            \centering
            \includegraphics[width = \textwidth]{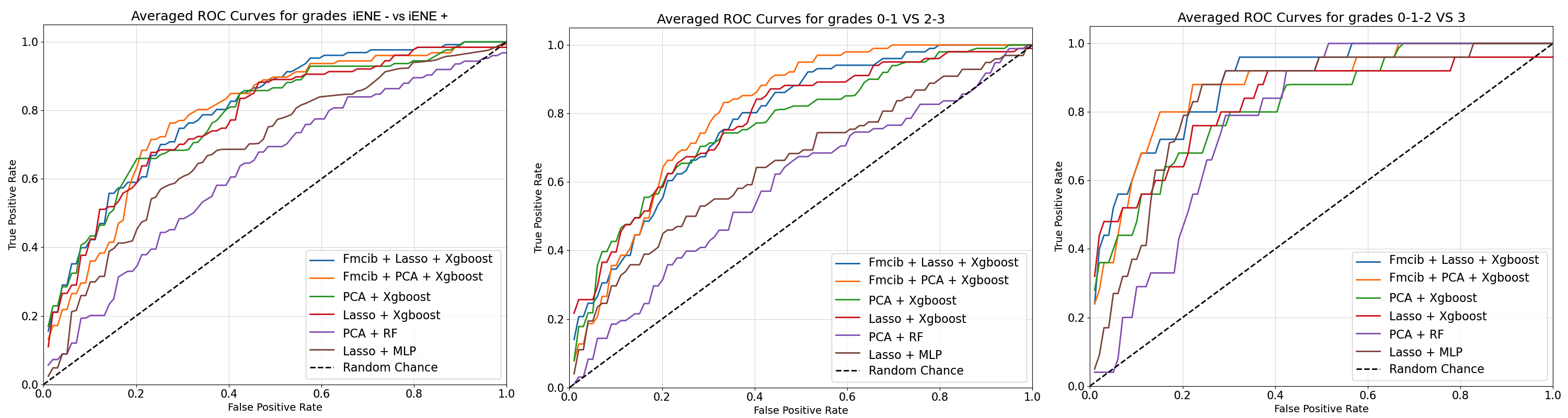}
            \caption{
                ROC curves obtained for the binary classification of dichotomized iENE classes, for 3 different dichotomization schemes (iENE$^-$ \textit{vs} iENE$^+$; 0-1/2-3; 0-1-2/3).
                Individual patient logits were obtained in a 5-fold cross validation scheme with the L2 norm penalized XGBOOST classifier.}
            \label{fig:ROCClassification}
        \end{figure*}

        Table \ref{tab:ResultsiENE} presents the binary iENE grade classification results. The highest AUC of 81.68 $\pm$ 5.70 was achieved using both radiomic and FMCIB model features with the XGBoost algorithm and Lasso feature selection. For the dichotomization case 0-1 vs. 2–3, the best performance was obtained also with the combined FMCIB+radiomics models, using principal component analysis (PCA) with XGBoost, yielding an AUC of 82.85 $\pm$ 6.03. For grade 3 identification, an AUC of 89.93 $\pm$ 5.07 was achieved using the XGBoost algorithm with Lasso feature selection on the combined radiomic and FMCIB feature set. The choice of feature selection method appeared to depend on the dichotomization scheme, with Lasso performing better for iENE- vs. iENE+, compared to PCA showing superior performance in the other case for 0-1/2-3 classification. The corresponding ROC curves for all dichotomization schemes are shown in Figure \ref{fig:ROCClassification}.
        
        The use of FMCIB features alone generally hindered performance. Indeed, these features alone led to significantly lower results across all schemes, with an AUC of 71.82 $\pm$ 2.32 for iENE classification. However, the combination of FMCIB features to radiomics produced the best result for all tasks, including grade 3 identification, the standard deviation was too large to confidently attribute the improvement to their inclusion.
        
\begin{figure}[h]
            \centering
            \includegraphics[width = \columnwidth]{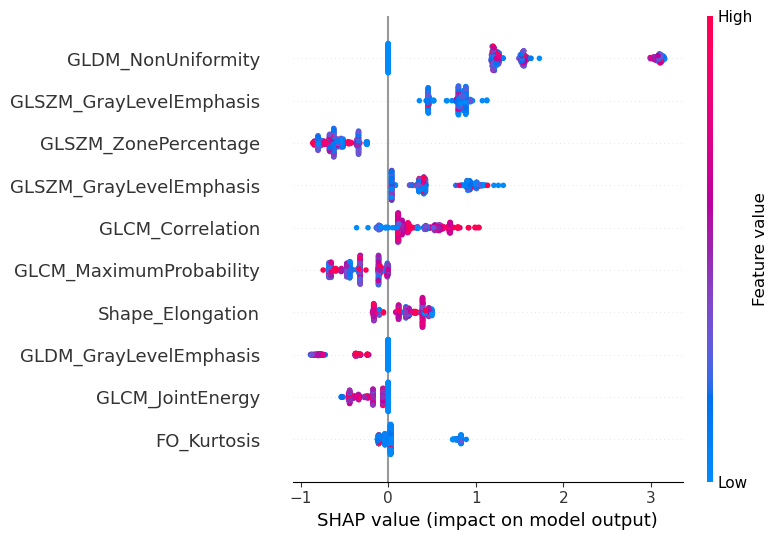}
            \caption{
                SHAP explanation analysis plots for the prediction of iENE- vs iENE + cases, with the top features obtained from the radiomics branch of the prediction framework.}
            \label{fig:SHAP}
        \end{figure}

        Figure \ref{fig:SHAP} presents the SHapley Additive exPlanations (SHAP) analysis of feature importance for iENE- vs. iENE+ classification. Higher levels of texture non-uniformity were most predictive of a positive ENE case. While most contributing features were texture-related, greater elongation distance of the node also contributed to iENE+ classification.
       
        \begin{table}[tbh]
            \centering
            \renewcommand{\arraystretch}{2.0}
            \resizebox{0.85\columnwidth}{!}{
            \begin{tabular}{llrrr}
                \hline
                Features                   & Method          & iENE$^-$ \textit{vs} iENE$^+$                 & 0-1 / 2-3                 & 0-1-2 / 3                 \\ \hline
                \multirow{4}{*}{Radiomics} & RF + PCA        & 78.12 $\pm$ 4.67          & 76.93 $\pm$ 5.98          & 89.87 $\pm$ 5.24          \\
                                           & Lasso + MLP     & 77.55 $\pm$ 2,65          & 74.98 $\pm$ 2.80          & 80.64 $\pm$ 17.31         \\
                                           & PCA + XGBOOST   & 76.39 $\pm$ 2.56          & {80.60 $\pm$ 4.52} & 86.60 $\pm$ 8.40          \\
                                           & Lasso + XGBOOST & {79.92 $\pm$ 2.51} & 78.00 $\pm$ 1.61          & 88.85 $\pm$ 5.27          \\ \hline
                \multirow{2}{*}{FMCIB}     & PCA + XGBOOST   & 70.33 $\pm$ 3.41          & 68.32 $\pm$ 5.51          & 80.52 $\pm$ 12.18         \\
                                           & Lasso + XGBOOST & 71.82 $\pm$ 2.32          & 64.98 $\pm$ 5.69          & 68.19 $\pm$ 10.36         \\ \hline
                \multirow{2}{*}{\begin{tabular}[l]{@{}l@{}}FMCIB \& \\ Radiomics\end{tabular}} & PCA + XGBOOST & 74.82 $\pm$ 7.25 & \textbf{82.85 $\pm$ 6.03}  & 87.99 $\pm$ 2.58 \\
                                           & Lasso + XGBOOST & \textbf{81.68 $\pm$ 5.70}          &  76.98 $\pm$ 7.65        & \textbf{89.93 $\pm$ 5.07}
            \end{tabular}}
            \caption{
                iENE grade classification results for 3 different dichotomization schemes (iENE$^-$ \textit{vs} iENE$^+$; 0-1/2-3; 0-1-2/3).
                Results show  the mean AUC ($\pm$ standard deviation) over the 5 tests fold obtained in a stratified cross-validation.
                Features indicate the CT feature extraction strategy used as inputs to the classification method.}
            \label{tab:ResultsiENE}
            \end{table}
            
    \subsection{Prognostic value of ENE in HPV+ OPC}
    
\begin{figure*}[tb!]
            \centering
            \begin{subfigure}[b]{\textwidth}
                \centering
                \includegraphics[width=\textwidth]{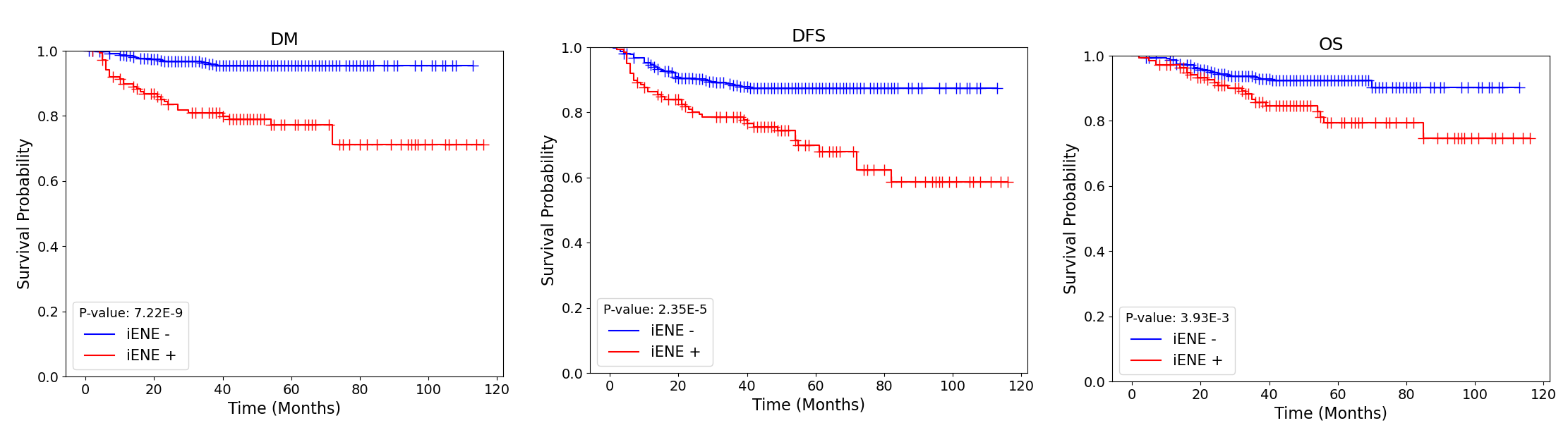}
                \caption{Kaplan Meier curve for the ground-truth ENE status as the univariate characteristic.}
            \end{subfigure}
            \begin{subfigure}[b]{\textwidth}
                \centering
                \includegraphics[width=\textwidth]{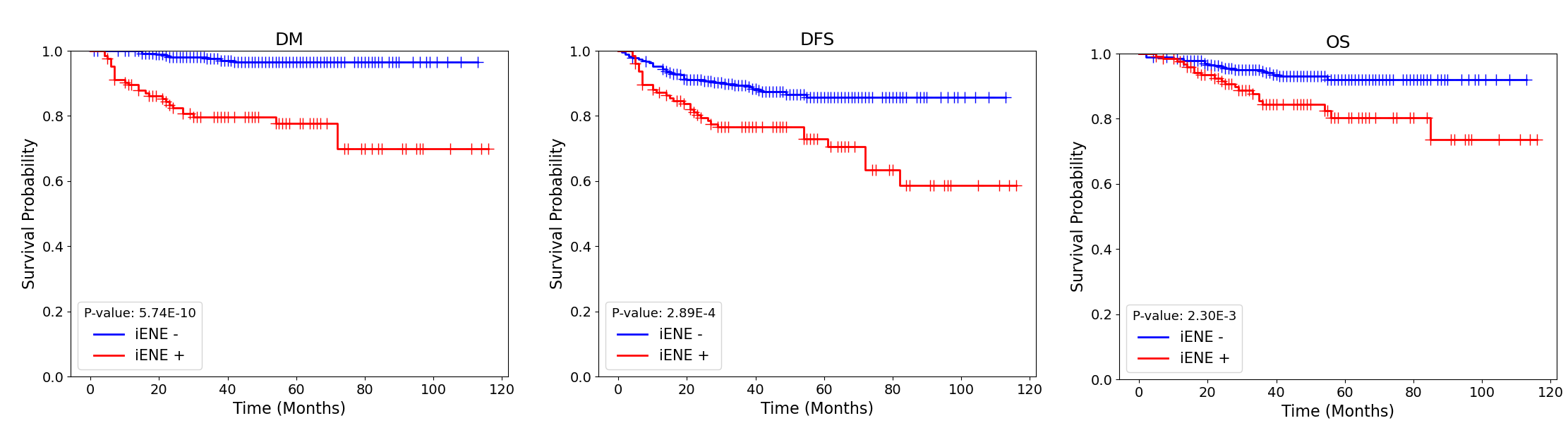}
                \caption{Kaplan Meier curve for the predicted iENE score as the univariate characteristic.}
            \end{subfigure}
            \caption{
                Kaplan Meier curves obtained by stratifying all 397 patients with the predicted iENE$^+$ score.
                (a) Survival curves from ground-truth pathology.
                (b) Survival curves from predicted iENE.
                Blue lines constitute the real or predicted iENE- groups, red lines show the real or predicted iENE+ population.
                The selected node for grade classification is obtained as the largest predicted volume.
                The Kaplan Meyer fit function is unimodal and takes as input the tresholded (from AUC) binary prediction value.
                Each curve displays as legend the p-value obtained in a logrank test (DM: $p=9.6\times10^{-10}$, DFS: $p=9.83\times10^{-5}$; OS: $p=2.8\times10^{-3}$).}
            \label{fig:KMcurve}
        \end{figure*}

        \paragraph{\textbf{Kaplan-Meier curves}} Figure \ref{fig:KMcurve} presents the Kaplan Meier survival curves for the three outcomes as described in Sec. \ref{Subsection KM}. We can observe a clear separation between groups based on the predicted iENE status from our classification pipeline. This indicates the importance of the iENE status in HPV-associated OPCs, as patients are more likely to develop worse clinical outcomes if they have higher iENE grades. With regards to the logrank statistical test, for the distant treatment failure, a p-value of $9.61\times10^{-10} < 0.05$ was obtained; for DFS, a p-value of $9.83\times10^{-5} < 0.05$; for survival, a p-value of $2.8\times10^{-3} < 0.05$. This indicates that all have a significant difference between survival groups. This experiment demonstrates that the iENE status may be used for outcome prediction, and that it may be an important predictive biomarker for cancer staging. 
        
        \begin{table}[h]
            \centering
            \renewcommand{\arraystretch}{2.0}
            \resizebox{\columnwidth}{!}{
            \begin{tabular}{lrrrrrr}
                \toprule
                Outcome & $\overline{C}$ & $C_1$ & $C_2$ & $C_3$ & SR & Proposed \\
                \midrule
                DM  & $2.34\times10^{-2}$ & $1.76\times10^{-2}$ & $3.13\times10^{-2}$          & $1.61\times10^{-1}$ & $7.78\times10^{-2}$ & \textbf{$8.02\times10^{-5}$} \\
                \midrule
                DFS & $1.98\times10^{-1}$ & $2.16\times10^{-1}$ & \textbf{$3.03\times10^{-2}$} & $3.39\times10^{-1}$ & $1.93\times10^{-1}$ & $3.30\times10^{-2}$ \\
                \midrule
                OS  & $2.62\times10^{-1}$ & $1.82\times10^{-1}$ & $1.86\times10^{-1}$          & $3.32\times10^{-1}$ & $2.82\times10^{-1}$ & \textbf{$2.90\times10^{-2}$} \\
                \bottomrule
            \end{tabular}}
            \caption{
                Log-rank statistical test for a separate set of 55 patients annotated by three separate radiologists.
                $\overline{C}$: A collegial decision involved a three way consensus over the case definition, and served as the ground truth for model training.
                $C_{1-3}$ refers to clinician 1, 2 or 3 respectively.
                SR: a single reader is simulated by randomly selecting one of the three annotations per patient, and bootstrapping the obtained results 10 000 times, giving the average score a unique reader could obtain.}
            \label{tab:Anotator Eval}
        \end{table}
        
        Table \ref{tab:Anotator Eval} presents the results comparing individual reviewers, the simulated reader (SR), and our model's predictions. For OS, DM and DFS predictions, the model demonstrated a statistically significant difference between event groups, whereas this is not the case for the SR. For all outcomes, inter-reader variability can be directly associated with event-based discrimination, as indicated by the provided p-values.
        
        \paragraph{\textbf{2-Year outcome risk prediction results}}
        
        \begin{figure*}[tbh]
            \centering
            \includegraphics[width = \textwidth]{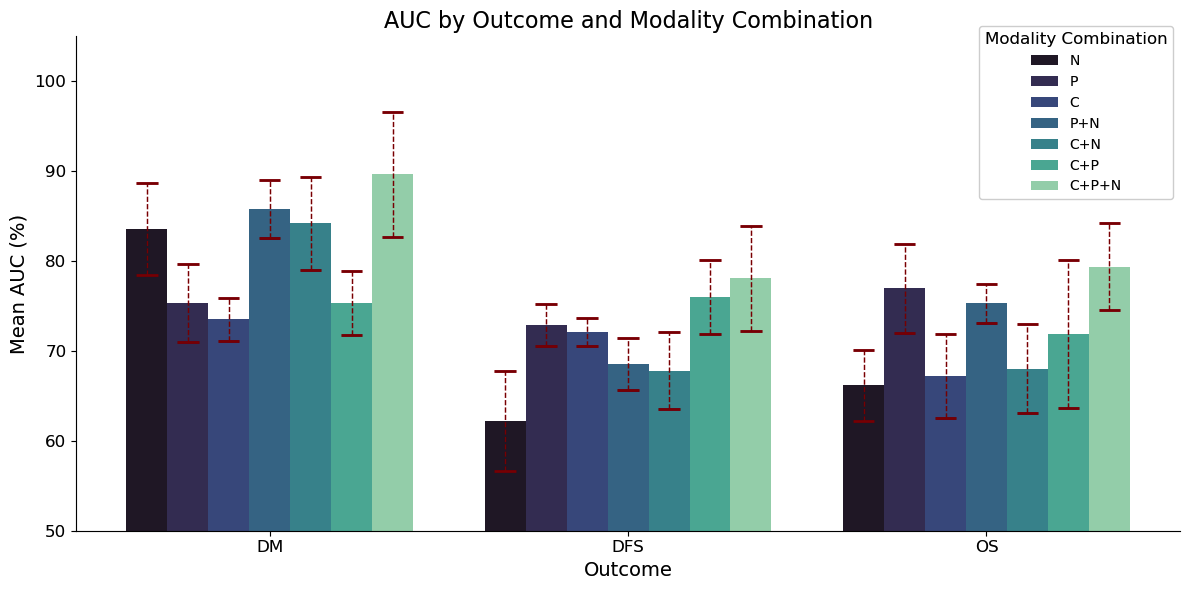}
            \caption{
                Modality ablation experiment results for the 2-year therapeutic outcome prediction model.
                For readability, C indicates the clinical variables (including the TNM staging), P represents the primary tumor value (GTV radiomic characteristics) and N the predicted nodal volume (iENE) radiomics.
                Each result presents the mean AUC over the 5 tests fold obtained in a stratified k-fold validation.
                For each modality and combination of modality, we present the best mean AUC obtained over a grid search as described in \ref{Setup Methodology}, as well as the standard deviation plotted in red dashed lines.} 
            \label{Figure:Multimodal Test}
        \end{figure*}
 
        \begin{table*}[tbh]
            \centering
            \renewcommand{\arraystretch}{1.3}
            \resizebox{\textwidth}{!}{
            \begin{tabular}{llccc}
                \hline
                Model         & Algorithm                       & DM 2 Y   (6$\%$)     & DFS 2Y (12$\%$)      & OS 2Y (5$\%$)       \\ \hline
                Cox           & Proportional Hazard Regression  & 69.7 $\pm$ 10.7 & 67.0 $\pm$ 8.5  & 63.4 $\pm$ 17.9 \\
                Kazmierski et al.\cite{RadCure}  & Logistic Regression             & 68.2 $\pm$ 6.4  & 59.4 $\pm$ 17.9 & 57.1 $\pm$ 11.6 \\
                Rebaud et al.\cite{rebaud_simplicity_2022}        & Bagged risk estimation          & 70.7 $\pm$ 5.0  & 62.9 $\pm$ 3.2  & 60.9 $\pm$ 12.7 \\
                Vallieres et al.\cite{vallieres_radiomics_2017} & Mrmre Random Forest             & 78.2 $\pm$ 7.2  & 67.2 $\pm$ 5.9  & 61.4 $\pm$ 12.4 \\
                Kazmierski et al.\cite{RadCure}  & Fuzzy Logistic Mrmr             & 80.5 $\pm$ 7.9  & 63.6 $\pm$ 6.3  & 57.9 $\pm$ 15.1 \\
                \textbf{AMO-ENE}& Multi-Head Attention Classifier & \textbf{88.2 $\pm$ 4.8}  & \textbf{78.1 $\pm$ 8.6}  & \textbf{79.2 $\pm$ 7.4} \\ \hline
            \end{tabular}}
            \caption{
                Binary 2-year classification algorithms for distant metastasis (DM), disease-free survival (DFS) and overall survival (OS).
                Proportion of (uncensored) events at the end of the study are reported in parentheses.
                Results show  the mean AUC ($\pm$ standard deviation) over the 5 cross-validation folds.
                All presented algorithms incorporate the primary tumor, predicted nodal and clinical characteristics.
                For every parametric model, results presented are the best performance obtained via grid-search as proposed by the respective authors.}
            \label{tab:OutcomePredictionResults}
        \end{table*}
        
        \begin{table*}[tbh]
            \centering
            \renewcommand{\arraystretch}{1.3}
            \resizebox{\textwidth}{!}{
            \begin{tabular}{lcclccc}
            \hline
            Model & \multicolumn{2}{c}{Features} & \multicolumn{1}{c}{Algorithm} & DM                   & DFS                  & OS                   \\ \hline
                    & Clinical       & Radiomics   &                               & \multicolumn{1}{l}{} & \multicolumn{1}{l}{} & \multicolumn{1}{l}{} \\ \cline{2-3}
            Cox     & $\checkmark$   &             & Cox Regression                & 66.0 $\pm$ 7.7       & 66.5 $\pm$ 6.3       & 64.4 $\pm$ 18.5      \\
            Rebaud et al. \cite{rebaud_simplicity_2022} &
              $\checkmark$ &
              $\checkmark$ &
              Bagged risk estimation &
              81.5 $\pm$ 8.1 &
              68.8 $\pm$ 4.2 &
              62.89 $\pm$ 8.1 \\
            Cox + XGBOOST iENE score &
              $\checkmark$ &
              $\checkmark$ &
              Cox Regression &
              80.1 $\pm$ 6.9 &
              68.6 $\pm$ 5.1 &
              {70.2 $\pm$ 9.1} \\
            \textbf{AMO-ENE (MHSA)} &
              $\checkmark$ &
              $\checkmark$ &
              AMO-ENE + MTLR &
              \textbf{83.3 $\pm$ 6.5} &
              \textbf{70.0 $\pm$ 8.1} &
              \textbf{71.3 $\pm$ 8.9} \\ \hline
            \end{tabular}}
            \caption{
                Survival risk estimation for metastatic failure (DM), disease-free survival (DFS) and overall survival (OS).
                Figures are the mean concordance index (C-index) ($\pm$ standard deviation) over the folds from the same 5-fold stratified cross-validation.
                For every parametric model, results presented are the best performance obtained  via grid-search as proposed by the respective authors.}
            \label{tab:SurvivalPredictionResults}
        \end{table*}
        
        \begin{table}[h]
            \centering
            {\fontsize{20pt}{20pt}\selectfont 
            \renewcommand{\arraystretch}{2.0}
            \resizebox{\columnwidth}{!}{%
            \begin{tabular}{lccccc}
            \hline
            Fusion Type         & Params (M) & Memory (MB) & Recall        & Specificity           & AUC          \\ \hline
            Early Fusion        & 1.95       & 15.4        & 60.3$\pm$19.2 & 87.5$\pm$3.5          & 76.4$\pm$7.5 \\
            Late Concatenation  & 0.86       & 15.3        & 55.0$\pm$28.6 & {86.6$\pm$5.2} & 80.1$\pm$7.1 \\
            Late Soft Attention & 1.17       & 12.7        & 75.2$\pm$23.2 & 88.4$\pm$5.2          & 86.7$\pm$6.6 \\
            {\textbf{AMO-ENE (MHSA)}} & 2.37 & 15.4 & \textbf{75.4$\pm$19.5} & \textbf{91.2$\pm$6.2} & \textbf{88.2$\pm$4.8} \\ \hline
            \end{tabular}%
            }
            }
            \caption{
                Ablation study for the modality fusion module of AMO-ENE, tested on the two year DM classification task.}
            \label{tab:AblationAMOattentiontype}
        \end{table}
        
        {\fontsize{20pt}{20pt}\selectfont
        \begin{table}[h]
            \centering
            \renewcommand{\arraystretch}{2.0}
            \resizebox{\columnwidth}{!}{
            \begin{tabular}{@{}cccccc@{}}
            \toprule
            \multicolumn{1}{l}{Model} & Number of Heads     & Memory (MB) & Recall                 & Specificity           & AUC                   \\ \midrule
            \multirow{4}{*}{AMO-ENE}  & 2                   & 15.62       & 71.2$\pm$27.0          & 74.3$\pm$21.5         & 75.4$\pm$10.1         \\
                                      & 4                   & 15.65       & 62.0$\pm$12.9          & 83.2$\pm$7.6          & 80.8$\pm$7.0          \\
                                      & \textit{\textbf{8}} & 15.68       & \textbf{75.4$\pm$19.5} & 86.6$\pm$6.5          & \textbf{88.2$\pm$4.8} \\
                                      & 16                  & 15.71       & 59.3$\pm$27.6          & \textbf{87.2$\pm$4.9} & 80.1$\pm$13.4         \\ \bottomrule
            \end{tabular}}
            \caption{
                Ablation study for the number of heads used in our multi-head attention fusion module of AMO-ENE, tested on the two year DM classification task.}
            \label{tab:AblationHeads}
        \end{table}}
        
        Table \ref{tab:OutcomePredictionResults} presents the results for the 2-year outcome prediction pipeline. Several multi-modal binary classifiers were compared to the proposed approach. The proposed multi-modal attention based classifier obtains the highest performance for all three reported outcomes, with 88.2 $\pm$ 4.8 for 2-year DM, 78.1$\%$ ($\pm$ 8.6) for DFS and 79.2$\%$ ($\pm$ 7.4) for OS prediction.
        
        Figure \ref{Figure:Multimodal Test} presents the comparative experiments for the outcome prediction model with all possible combinations of input features provided (clinical, GTV, ENE). We observe that combined characteristics of all three inputs allow for an improved outcome prediction over the clinical baseline (88.2$\% > 73.5\%$ AUC on DM prediction), which includes actual AJCC cancer staging statuses. For metastatic involvement, a major component residing in the iENE characteristics, scoring 83.1$\%$ when considered as the sole input for DM. Furthermore, DFS and OS also benefited from the inclusion of all three components, reaching 78.1$\%$ ($\pm$ 8.6) and 79.2$\%$ ($\pm$ 7.4), respectively.  
        
        \paragraph{\textbf{Multi-bin risk modeling outcome estimation results}}
        
        Table \ref{tab:SurvivalPredictionResults} presents the C-indexes of trained survival models on the multi-omics dataset. The results for the Cox regression model only include the clinical metadata as they could not converge on multi-omics. Inclusion of the nodal component through the predicted iENE score or the multi-omics approach improved risk assessment for all three outcomes (83.3> 66.0$\%$ for DM). Performance did also improve for OS by integrating MTLR fusion (71.3$\%$ AUC) over a Cox model integrating the iENE score (70.2$\%$ AUC).
        
    \subsection{Ablation experiments}
    
        Finally, we present the results of ablation studies designed to evaluate the relative importance of key components in AMO-ENE for integrating multi-modal input.
        
        Table \ref{tab:AblationAMOattentiontype} reports performance on the DM prediction task using different modality fusion strategies, while keeping all other components of AMO-ENE unchanged. Late fusion approaches all performed better than early fusion, with significant improvements in performance with comparable or lower parameter and memory requirements. Additionally, we observed that attention-based fusion outperformed latent concatenation, underscoring that enabling cross-modal interactions through attention mechanisms further enhances predictive performance, albeit at the cost of increased parameter count.
        
        Table \ref{tab:AblationHeads} presents the effect of varying the number of attention heads in the MHSA fusion module. Using 8 heads yielded the best classification performance, with notable gains over using just 2 heads. However, increasing the number of heads to 16 did not result in further improvements. These results underscore the importance of hyperparameter tuning when integrating MHSA for modality fusion.

    \section{Discussion}

        In this study, we proposed a framework to automatically detect and characterize extra nodal extensions in HPV positive OPC patients, obtaining their grading score which is used as a prognostic imaging marker of radiotherapy outcomes. We first proposed an automated nodal extension segmentation method from CT scans, which avoids time-consuming and interpretation-subjective tasks by physicians. From the obtained segmentation masks, the model extracts specific biomarkers to automatically assess the iENE grade. We linked the obtained score to DM, DFS and OS outcomes and compared its prognostic value against single annotators and current clinical guidelines. Finally, we proposed a multi-omic attention based network to predict these three outcomes at two years, combining and assessing the role of lesions for this survival analysis task.
        
        Segmentation of ENE remains a challenging task, with ground truth annotation margins significantly affecting both the training and evaluation of prognostic models. The increased variability in Dice scores underscores this issue, particularly for small nodes, which tend to have a higher surface-to-volume ratio. We achieved a maximum overall Dice of 83.5 $\pm$ 4.1 using the SwinUNETRv2-based model and nodal selection model, followed by nnUNet, indicating that the size of the dataset was better suited to hybrid CNN+ViT-based methods. The use of a foundation prompt model proved detrimental to segmentation performance. We hypothesize that SamMed3D was not pre-trained on HNC nodal lesions and, as a result, struggled to accurately identify segmentation boundaries—even with 10 points provided.
        It should be noted that segmentation performance only reflects a model’s ability to replicate the provided annotations, which may suffer from inter-annotator variability as is our case. Segmentation performance also does not necessarily reflect the prognostic relevance of the predicted masks.
        
        From the extracted masks, specific biomarkers were subsequently extracted, allowing for the automated grading of an iENE score. To that end, we compared radiomics against deep foundation model features \cite{fmcib} and obtained the best result with the combination of both, achieving 89.93$\%$ ($\pm$ 5.07) AUC score for iENE$^-$ (0) against iENE$^+$ (1/2/3) classification. We hypothesize that radiomic features harness strong information for ENE discrimination, as the size of the object is indicative of higher grades, the texture indicative of tumoral spread and the sphericity of the node indicates how likely one is to have its capsule ruptured. Deep features alone performed worse than the radiomics approach: we hypothesize that FMCIB features may be less explicit in their ability to model nodes and the imposed fixed region of interest of the FMCIB model hindered uniformity of context for our variable collection of object sizes. Still, classification performance needs to placed in the context of the provided annotations, as predicted iENE scores may be more or less indicative of the actual status of the disease.
        
        To evaluate the prognostic relevance of the predicted iENE score, we compared its stratification performance against both ground truth annotations and individual clinician assessments across OS, DM and DFS. The groupings generated by the model's predicted iENE score consistently achieved statistical significance across all outcomes, with p-values of $8.02 \times 10^{-5}$ for DM, $3.30 \times 10^{-2}$ for DFS, and $2.90 \times 10^{-2}$ for OS based on the log-rank test. 
        
        These values are notably more discriminative than the collegial decision ($\overline{C}$), which yielded less significant p-values: $2.34 \times 10^{-2}$ (DM), $1.98 \times 10^{-1}$ (DFS), and $2.62 \times 10^{-1}$ (OS). When compared to individual raters and against the simulated single reader (SR) scenario, which mimics clinical uncertainty by bootstrapping individual annotations, weaker association with outcomes was found, failing to reach statistical significance in all cases. These results indicate that the model not only surpasses individual expert assessments but also outperforms a robust consensus label in stratifying patients by risk. We hypothesize that this is due to the inherent subjectivity of iENE grading from radiologists, which may limit human-level reproducibility. In contrast, the model leverages standardized, data-driven features, allowing for a more robust and precise interpretation of underlying imaging biomarkers. 
        
        Finally, from the nodal characteristics, we introduced an outcome prediction model and compared it to other multi-omic fusion methods in the literature. Our model achieved the best performance for all three outcomes at the two year landmark, with 88.2$\%$ ($\pm$ 4.8), 78.1$\%$ ($\pm$ 8.6) and 79.2$\%$ ($\pm$ 7.4),  for DM, DFS and OS, respectively. For DM, the characterization of the nodal extension was the main contributor to the prediction (83.1$\%$ as standalone), but was  complemented by both the clinical and tumoural primary information, increasing the AUC by 5.1$\%$. This suggests that extra nodal extension expresses the ability of an HPV-associated OPC to metastasize, and that its characteristics contain biomarkers predictive of the outcome. DFS also showed the best results with the combination of all three modalities. Patients' survival at two years was mainly dictated by the primary tumor characteristics (82.1$\%$  $>$ 66.2$\%$ in AUC for node radiomics), suggesting that metastatic failure through nodal extensions is not the main life threat at this time point. 
        
        From separate multi-modal inputs, we show that deep attention based modality fusion allows to improve the combination and exploration of interactions than the proposed selection algorithms.
        
        We also observed an improvement in performance in survival estimation across the full follow-up period (2 months - 7 years) by integrating the MTLR framework \cite{MTLR_base} into our multi-omic fusion model. The most notable improvement was seen in the distant metastasis (DM) prediction, where the model achieved a C-index of 83.3$\%$ (±6.5), substantially outperforming the Cox regression baseline trained on clinical features alone (66.0$\%$ (±7.7)) as well as the Cox model integrating the predicted iENE score in addition to the clinical criteria (80.1$\%$ (±6.9)). This highlights the significant impact of extra-nodal extension features in modulating metastatic risk over time. Disease-free survival (DFS) estimation also benefited from this approach, reaching a C-index of 70.0$\%$ (± 8.1). Overall survival (OS) prediction improved compared to the clinical baseline (71.3$\%$ vs. 64.4$\%$), although it did not surpass the Cox model that incorporated the iENE score. This may be attributed to the relatively small number of uncensored OS events (n=38), which limits the model’s ability to learn from death-associated patterns. Furthermore, OS is inherently complex to model, as it may be influenced by multiple competing risk inducing conditions, leading to unreported causes of death not directly related 
        to OPC \cite{riskscores}. 
        
        This study has some limitations that would need to be addressed in future works. First, the model was trained using data from a single center, and would benefit from an external  validation set to evaluate generalizability, choice of scanning protocols and treatment variations. Fluctuations inter institutions for iENE annotation practices would be specifically important to assess. In addition, our study population is predominantly male, leading to potentially unexplored sex specific discrepancies in the obtained results. The current dataset lacks information on race and ethnic background, preventing the evaluation of potential performance disparities across under-represented groups. Finally, the timing disparity in the follow up exams may affect risk estimation. The inclusion of larger datasets with long-term follow-ups would improve robustness of presented results.   In terms of reproducibility, potential bias may originate  from the strong of imbalance of men vs women the cohort (317 M vs. 80 F). However, this distribution follows epidiomelogical observations in oropharynx cancer. Furthermore, the method was tested only on an in-house dataset, where patients were followed with 2 and 5-year follow-ups and with pathological confirmation, which is not available in public datasets.  Evaluation was performed using both cross-validation and separate test set, with cross-fold validation was confined to subject belonging only to one fold. This ensured no data leakage was introduced.
        
    \section{Conclusion}
        
        In this paper, we proposed AMO-ENE, an attention-based multi-omics fusion approach framework for the staged detection and classification of imaging extra-nodal extension (iENE), used for outcome prediction from head and neck planning CT scans and clinical data in HPV-associated OPC patients. This pipeline could be used  in a clinical setting with the potential to alleviate the  burden of the nodal assessment task and provide a more robust classification system for cancer staging. Notably, this standardized approach could help mitigate inter-rater variability in iENE grading, with the added benefit of promoting the access of prognostic tools for centers without specialized expertise. The findings of this work support the inclusion of the iENE status in cancer staging evaluation for HPV positive OPCs. Further work will be conducted on integrating a larger cohort of patients, with longer follow up prospective evaluations and multi-site cross institutional databases. Additionally, we will further evaluate our outcome prediction model by including a larger cohort of organs at risk and lesions in order to enhance predictions by mapping the complex head and neck structure. Finally, a clinical deployment may be warranted to evaluate the clinical impact of such model on treatment decision making.

        \paragraph{\textbf{Ethical statement}} Methods were carried out in accordance with relevant guidelines and regulations established by national standards and the institutional board. This study was performed in line with the principles of the Declaration of Helsinki. Ethics approval was granted by the Institutional Review Board (IRB) for human studies of the Centre Hospitalier Université de Montréal (CHUM). Informed consent was waived by the IRB (Comité Éthique de la Recherche 18.194) of CHUM as the data used in the study was retrospective.

    \section{Data and code availability}
    
    \noindent The datasets generated and/or analysed during the current study are not publicly available due to containing personal identifying information, but are available from the corresponding author on reasonable request.
    
    \noindent The underlying code for this study and training/validation datasets is not publicly available but may be made available to qualified researchers on reasonable request from the corresponding author.

    \section{Author Contribution Statement}
    
    \noindent GH was a major contributor in writing the manuscript.
GH, WL, CB, and GD were the main technical contributors and implemented, analyzed and interpreted the experiments and results.
KN, AC, EF, PN, LL and HB curated, interpreted and validated the provided experimental data.
LL, WL and HB reviewed and revised the manuscript.
SK designed the study, reviewed and revised the manuscript.
        All authors read and approved the final manuscript.

    \section{Funding Declaration} 


    \noindent This research has been funded in part by the Natural Sciences and Engineering Research Council of Canada (NSERC).
        LL was funded with the Fonds de Recherche Quebec Sante (FRQ-S)/Fondation de L’Association des Radiologistes du Quebec (FARQ), clinical research scholarship Junior 1 salary award (311203).
        The funder played no role in study design, data collection, analysis and interpretation of data, or the writing of this manuscript. 

     \section{Acknowledgments}
     The authors would like to thank the CHUM CITADEL team for the data management.
 
    \section{Competing interests}
    
    \noindent All authors declare no financial or non-financial competing interests.

\bibliographystyle{abbrv}
    \bibliography{sample}

\end{document}